# New Routes to Solitons in Quasi One-Dimensional Conductors.

## S. Brazovskii [1,2]

[1] LPTMS-CNRS, UMR 8626, Univ. Paris-Sud bat. 100, Orsay, F-91405
[2] Landau Institute for theoretical Physics, Moscow, Russia.

We collect evidences on existence of microscopic solitons, and their determining role in electronic processes of quasi-1D conductors. The ferroelectric charge ordering gives access to several types of solitons in conductivity and permittivity, and to solitons' bound pairs in optics - both in insulating and conducting cases of TMTTF and TMTSF subfamilies. The excursion to physics of conjugated polymers allows to suggest further experiments. Internal tunnelling in Charge Density Waves goes through the channel of "amplitude solitons", which correspond to the long sought quasi-particle - the spinon. The same experiment gives an access to the reversible reconstruction of the junction via spontaneous creation of a lattice of $2\pi$ solitons - a grid of dislocations. The individual $2\pi$ solitons have been visually captured in recent STM experiments. Junctions of organic and oxide conductors are anticipated to show similar effects of reconstruction.

## I. Introduction.

Role of solitons in electronic properties has been appreciated in theories of mid 70's and firstly accessed in experiments of early 80's. New motivations come from discoveries in organic conductors, from new accesses to Charge Density Waves, from synthesis of new conducting polymers. Also, new interests to explore strongly correlated systems in electronic devices pose the question: shall we find there something less expected than electrons and holes of conventional semiconductors? ("Doping of insulators is a paradigm in physics of strongly correlated solids" – quote H. Fukuyama lecture at the Yukawa Institute anniversary.) We shall always address systems with a spontaneous symmetry breaking. It leads to degenerate equivalent ground states, and then the soliton is a connecting kink between them, see the figure below.

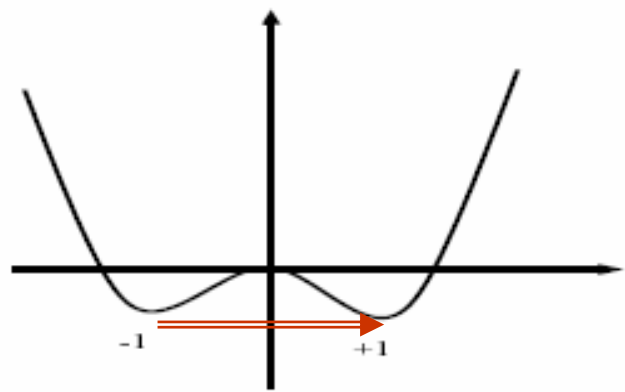

Figure. Total energy as a function of a symmetry breaking coordinate, e.g. bond dimerization or site disproportionation. The arrow shows a soliton as a domain wall connecting two degenerate ground states.

## II. Ferroelectricity and charge ordering in quasi 1D organic conductors as a scientific polygon for solitons.

The ferroelectricity (FE), and its more general basis of the charge ordering (CO) [1,2] in organic conductors provide a particular facility to see solitons [3]. The conductivity shows a thermal activation which is due to a spontaneous symmetry breaking, but what does conduct in these narrow gap semiconductors or Mott insulators? Apparently NOT the electrons: there is no gap in spin susceptibility, which stays flat. This is the clearest case of conduction by charged spinless solitons – holons, which are interpreted as so called $\pi$-solitons [3].

There are several equivalent views on the state with the CO and FE [1,3]: combined Mott state, commensurate Wigner crystal, mixed site-bond $4K_F$ CDW. The order parameteris is $\sim \cos(2\varphi+x\pi/a)$, its phase $\varphi$ is locked by the commensurability energy $H_U \sim -U\cos(2\varphi-2\alpha)$. The potential U comes from the CO, the phase centre shift $\alpha$ is the signature of the FE, the $H_U$ is doubly degenerate between $\varphi= \alpha$ and $\varphi= \alpha+\pi$. It allows for phase $\pi$ solitons - holons with charge e, they are the CO vacancy defects. (There are also two more specific types of solitons [3]).

There are many indirect evidences on solitons: separation of charge from spin; enhanced activation energy for interchain transport. But do we see them in dynamics - optics or tunneling?

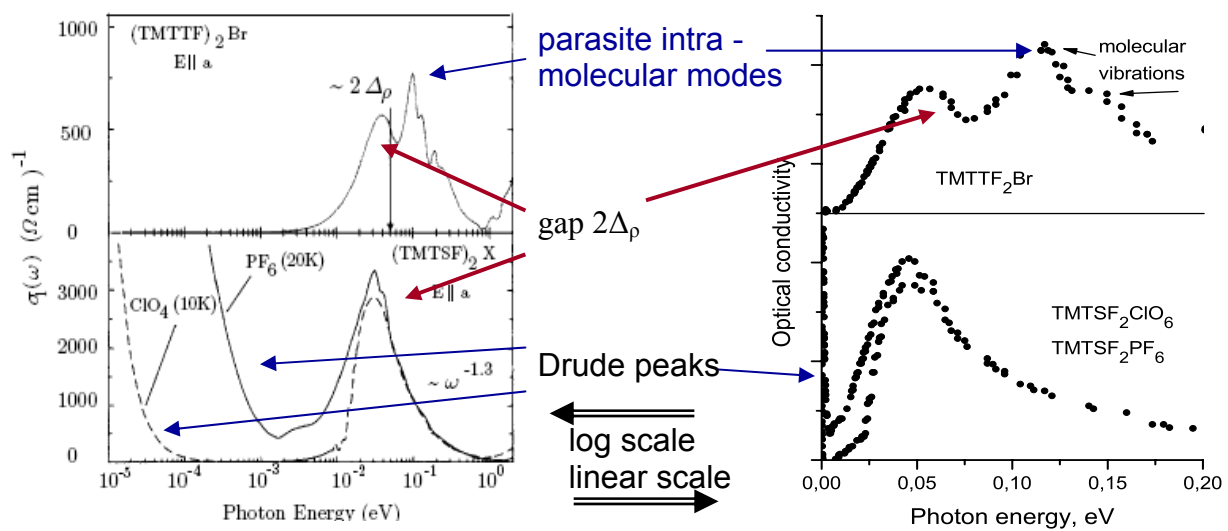

Figure 1. Comparison of optical absorption in TMTTF and TMTSF subfamilies, after [4].

Figure 1 shows the absorption spectra - the optical conductivity $\sigma(\omega=E/\hbar)$ of several TMTCF compounds. Left panel gives plots in logarithmic scale of energies, as published in [4]; right panel shows our replotting in the linear scale, more appropriate for gap features. At both panels, the upper plots show the data for the TMTTF$_2$Br, which is known to possess the FE CO state leading to the known value of the conduction gap $\Delta_\rho$. This gap is the lowest among all $TMTTF_2X$ compounds, which gives a unique chance to observe it below the intensive spectrum of intra-molecular vibrations. (Experiments under pressure will recover the spectra of decreasing gaps also in other compounds of this sub-family.) The lower plots show the data for nominally metallic $TMTSF_2X$ compounds. The CO might be weaker for the last subfamily, which allows observing the intrinsic optical features well below the parasite vibrations. Here the gaps were not accessed by other methods, but notice a striking similarity of gap features for explicit ($TMTTF_2Br$ case) and hidden ($TMTSF_2X$ cases) Mott states. In the following we shall use this advantage by *interpreting strong features observed in many TMTSF compounds in terms of the well established picture of the CO in most of the TMTTF ones*.

A principal question of dynamics is: if the peak is due to the edge $E_g = 2\Delta$ for production of pairs of unbound kinks or due to the lower optical absorption edge of the exciton – a bound kink-antikink state? Joint experiments on absorption and photoconductivity are necessary to distinguish productions of free kinks pairs and of their bound states - excitons or breathers. An interpretation suggested at the figure 2 exploits principal and secondary features of the spectra.

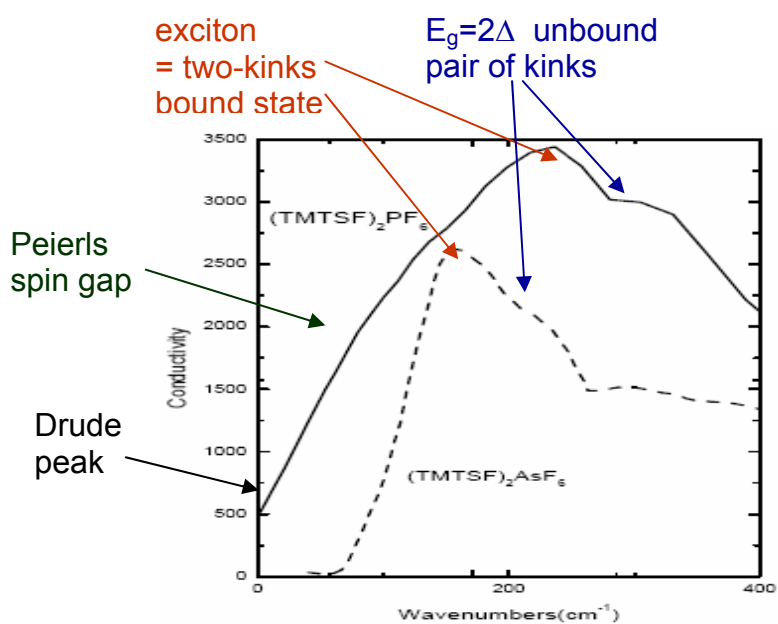

Figure 2. Interpretation of optics on TMTSF in terms of firm expectations for the CO state in TMTTF. Optical conductivity is replotted from data of Dressel group (unpublished).

## III. Lessons from conducting polymers.

Usually a possibility for appearance of excitons is ignored in optics of all kinds of CDWs, and the absorption peaks are fitted to free particle creation law in 1D : $I \sim (\omega - E_g)^{-1/2}$. This suggestion is challenged by the rich experience from conjugated polymers. Figure 3 demonstrates a very similar shape of the optical absorption in the polymer PPV and in $TMTSF_2X$. But for polymers we know definitely that the exciton component is present at the peak: there is a luminescence near its tip. (Always there is at least a stimulated luminescence – none was attempted to be measured in organic good glass.)

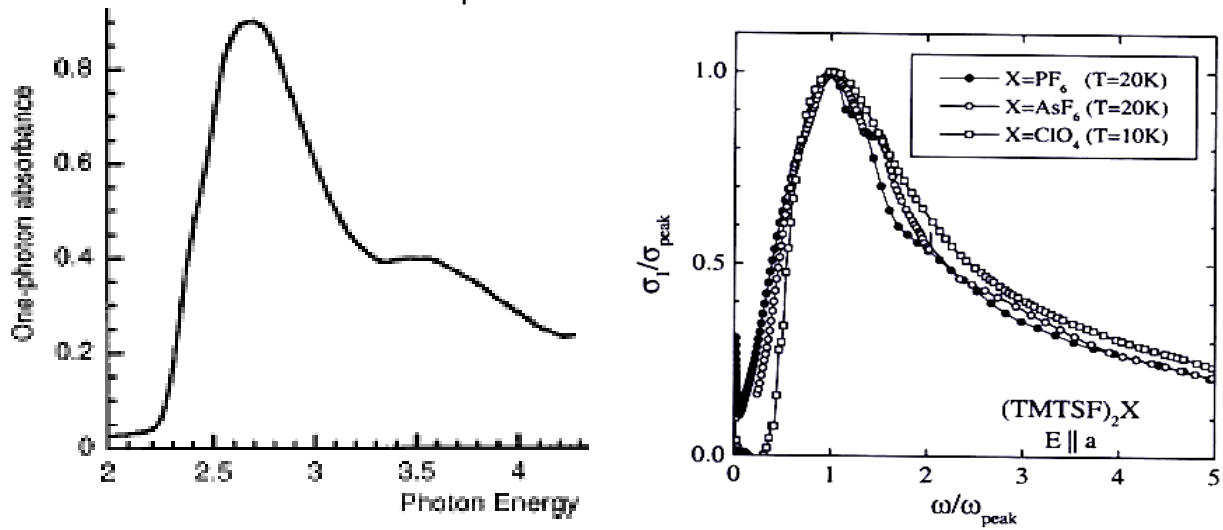

Figure 3. Comparison of optical spectra of the explicitly gapful conjugated polymer PPV [6], and the organic conductors TMTSF$_2$X [7] with a presumably hidden charge ordering gap. Is it a two-particle edge or a lower exciton? BOTH, according to polymers experience!

Nevertheless in polymers, in view of natural widths, the exciton and free pairs have never been resolved in the direct absorption, which originated a long dispute on the nature of the light emitting exciton (which is so important in applications), see [5]. The problem has been clarified only recently thanks to experiments on photoconductivity in presence of a high electric field [8], see [9] for a corresponding theory. Figure 4 shows that the two mechanisms become separated. Apparently, the organic CO semiconductors are waiting for these kinds of experiments, which may even result in applications for the IR optics.

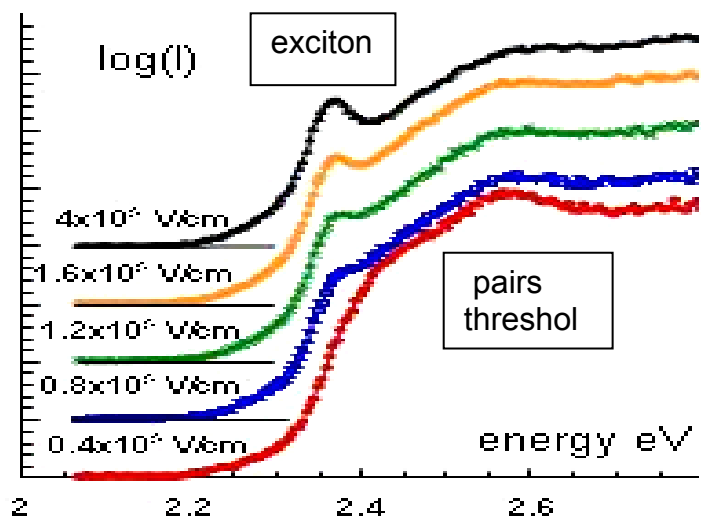

Figure 4. Separation of the two mechanisms by means of the high field photo-conductivity. Visualization of the exciton by electric field (field induced dissociation) [8]. The exciton gives the low energy peak, while the pair creation threshold is seen

as a kink at a higher energy. Notice the logarithmic scale of the intensity – there is an exponential inflation of intensity between the exciton and the free pairs.

## IV. Observations of solitons and their arrays in incommensurate CDWs.

Now we are changing gears to inorganic quasi-1D materials, always addressing the material $NbSe_3$. Contrary to the Charge Ordering, the conventional CDW is a crystal of electron pairs. Hence, the lowest energy current carrier must be a charge 2e defect of adding/missing one period. In terms of the phase φ of the CDW ~ $\cos(2K_f x + φ)$, it is the ±2π soliton. All these elements altogether (chains, the CDW, and the soliton) have been captured in recent STM experiments [10], figure 5.

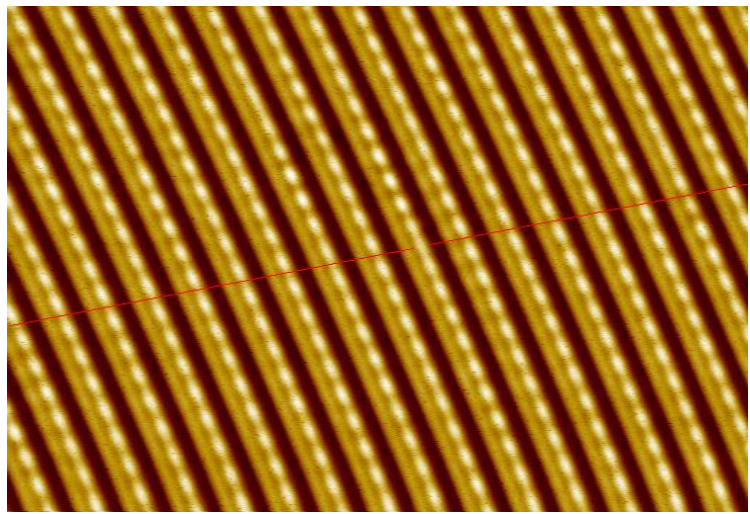

Fugure 5. STM scan of the $NbSe_3$, after [10]. Periodic white spots are the CDW. The horizontal line is drawn as the order parameter front $\cos(2K_f x)=0$, its position is shifted by just one half of the period in the center, at the defected chain. The defect is the 2π soliton - 2e hole of the CDW crystal.

But what should happen if the singlet is broken? It will not be an expected liberated electron-hole pair, but two spin carrying "amplitude solitons" – zeros of the order parameter distributed over a number of periods, figure 6. This creature appears in pair-breaking: amplitude soliton with the energy ≈2/3Δ, total charge 0, spin ½. This is the CDW realization of the SPINON.

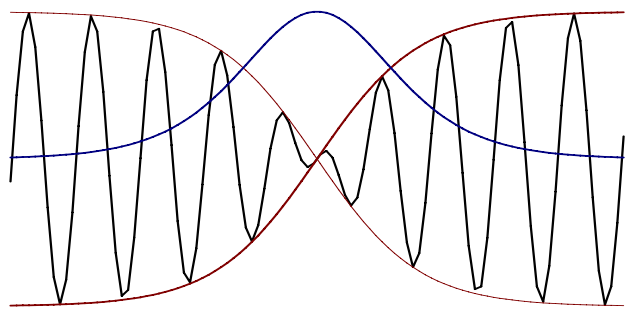

Figure 6. Oscillating electronic density, overlap amplitude soliton A(x), midgap state and spin distribution profile.

Walls of these solitons is just what might appear after the pair-breaking of CDWs at high magnetic field [11], or as a hunted FFLO phase in organic and other layered superconductors. Ref. [12] gives a literature review on theories of solitonic lattices. The question is: do we see that in dynamics? Yes, in inorganic CDWs.

Figure 7 quotes results of a new type of tunneling experiments [13] – internal interlayer tunneling in overlapping mesa-junctions, fabricated by the focused ion beam technology. Together with the expected, while remarkably sharp, peak at the free electron-hole pairs threshold $2\Delta$, we see also the lower energy strong feature, positioned close to the theoretical value $E_{as}=2/\pi\Delta$ for the amplitude soliton energy. Moreover, at much lower energies, forbidden for pair-breaking processes, we see the threshold $V_t\approx 0.2\Delta$, which terminates the bi-electron tunneling processes leading to formation of just the solitons captured at figure 5. Even the oscillating fine structure within the gap is not a noise – it records processes of sequential entering of dislocation lines, which are aggregates of $2\pi$ solitons, into the junction. For more on interpretations, see [14].

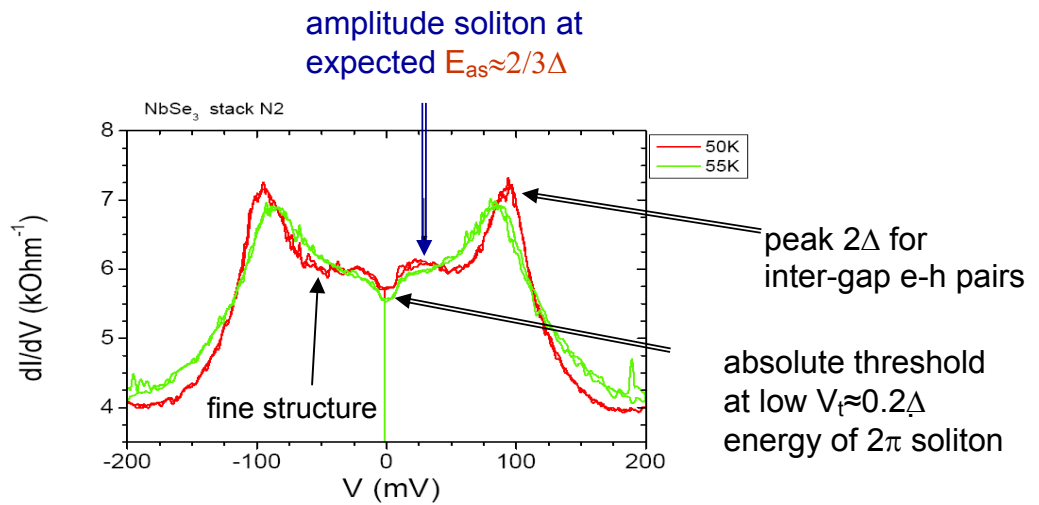

Fig.7. Multiple features recovered by new methodics of tunneling experiments, applied to the NbSe$_3$ [13]. The spectra correspond to the lower CDW, at two temperatures close to the 3D ordering transition at 59K. All the features scale with the gap value $\Delta(T)$.

## V. SUMMARY

In summary, the experiment and the theory prove existence of solitons in single- or bi-electronic processes in various quasi 1D materials with a spontaneous symmetry breaking. These systems feature self-trapping of electrons into midgap states and separation of spin and charge into spinons and holons, sometimes with their reconfinement at essentially different scales. Solitons take over band electrons in the role of primary excitations in optics and as charge or spin carriers. Solitons' aggregates appear as lattices in high magnetic fields or as a junction reconstruction in electric field. Physics of solitons clearly demonstrates necessity of reconciliation of unfortunately diverged directions of synthetic metals: organic stacks, inorganic chains, conjugated polymers.

The author acknowledges collaboration with N. Kirova, Yu.I. Latyshev, S. Matveenko, P. Monceau, F.Ya. Nad and receiving unpublished results from C. Brun and Z.Z. Wang, and from M. Dressel. A support was provided by the INTAS grant 7972 and by the ANR program (the project BLAN07-3-192276).